\documentclass[conference, 10pt]{IEEEtran}
\IEEEoverridecommandlockouts
\usepackage{cite}
\usepackage{amsmath,amssymb,amsfonts}
\usepackage{algorithmic}
\usepackage{graphicx}
\usepackage{textcomp}
\usepackage{hyperref}
\usepackage{xcolor}
\usepackage{comment}
\usepackage{siunitx}
\def\BibTeX{{\rm B\kern-.05em{\sc i\kern-.025em b}\kern-.08em
    T\kern-.1667em\lower.7ex\hbox{E}\kern-.125emX}}
\usepackage{tikz}
\usepackage{tikz, tikzscale}
\DeclareRobustCommand\sampleline[1]{%
  \tikz\draw[#1] (0,0) (0,\the\dimexpr\fontdimen22\textfont2\relax)
  -- (2em,\the\dimexpr\fontdimen22\textfont2\relax);%
}
\begin{document}
\title{Room Transfer Function Reconstruction Using Complex-valued Neural Networks and Irregularly Distributed Microphones
\thanks{This work has been funded by ``REPERTORIUM project. Grant agreement number 101095065. Horizon Europe. Cluster II. Culture, Creativity and Inclusive Society. Call HORIZON-CL2-2022-HERITAGE-01-02."}}
\author{\IEEEauthorblockN{Francesca Ronchini, Luca Comanducci, Mirco Pezzoli, Fabio Antonacci, Augusto Sarti}
\IEEEauthorblockA{\textit{Dipartimento di Elettronica, Informazione e Bioingegneria (DEIB), Politecnico di Milano}\\
Piazza Leonardo Da Vinci 32, 20133 Milan, Italy\\
name.surname@polimi.it}
}
\maketitle
\begin{abstract}
Reconstructing the room transfer functions needed to calculate the complex sound field in a room has several important real-world applications. 
However, an unpractical number of microphones is often required. 
Recently, in addition to classical signal processing methods, deep learning techniques have been applied to reconstruct the room transfer function starting from a very limited set of measurements at scattered points in the room.
In this paper, we employ complex-valued neural networks to estimate room transfer functions in the frequency range of the first room resonances, using a few irregularly distributed microphones. To the best of our knowledge, this is the first time that complex-valued neural networks are used to estimate room transfer functions.
To analyze the benefits of applying complex-valued optimization to the considered task, we compare the proposed technique with a state-of-the-art kernel-based signal processing approach for sound field reconstruction, showing that the proposed technique exhibits relevant advantages in terms of phase accuracy and overall quality of the reconstructed sound field. For informative purposes, we also compare the model with a similarly-structured data-driven approach that, however, applies a real-valued neural network to reconstruct only the magnitude of the sound field. 
\end{abstract}
\begin{IEEEkeywords}
sound field reconstruction, RIR interpolation, complex-valued neural network, room acoustics
\end{IEEEkeywords}
\section{Introduction}
\label{sec:i-introduction}

\begin{figure*}[h!]
    \centering
    \includegraphics[width=.8\linewidth]{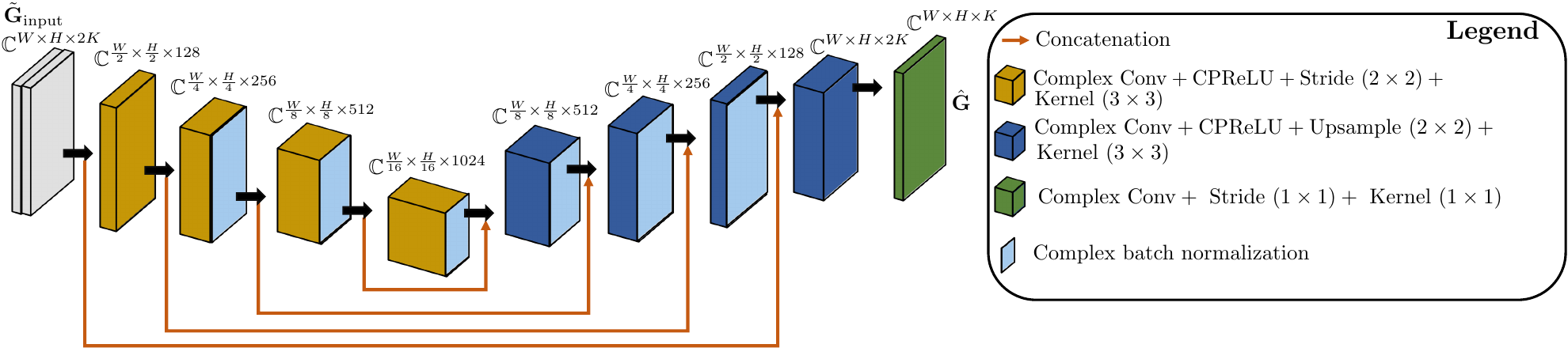}
    \caption{Schematic representation of the proposed CVNN. The first block represents the input of the network, followed by the four complex-valued convolutional encoder's layers, and the five complex-valued convolutional decoders's layers.}
    \label{fig:architecture}
\end{figure*}

Immersive audio plays a crucial role in virtual and augmented reality applications, leading to a growing interest in the navigation of the acoustic scene \cite{tylka2020fundamentals, cobos2022overview}, often referred to as 6 Degrees of Freedom (6DOF). 
To develop such solutions in an effective way, it is necessary to reconstruct the sound field in a large area, with the availability of room transfer functions measured in just a few points in the room. Solutions for sound field reconstruction incorporate model-based and data-driven approaches.
Model-based strategies leverage existing knowledge of acoustic principles to address the task of sound field reconstruction.  These can be categorized into parametric~\cite{pezzoli2020parametric, pulkki2018parametric},  and non-parametric models~\cite{
ueno2018kernel, ribeiro2023kernel, pezzoli2022sparsity
}. 
Parametric models represent the sound scene using a limited set of parameters, such as source position and directivity~\cite{pezzoli2020parametric}, to effectively convey spatial audio information. 
Non-parametric models, on the other hand, exploit combinations of plane waves~\cite{jin2015theory} or spherical waves~\cite{
pezzoli2022sparsity} to accurately reconstruct the acoustic field.   
Some non-parametric solutions adopt as a prior the Helmholtz equation to derive the kernel of the interpolation. In~\cite{caviedes2021gaussian}, the authors explore the use of Gaussian Process (GP) regression for sound field interpolation, while kernel-based interpolation approach have been proposed in~\cite{ueno2018kernel} and in~\cite{ribeiro2023kernel}. 

Mainly motivated by the encouraging results obtained in other acoustic tasks~\cite{ bainco2019acousticdeepreview}, data-driven approaches have found application also in addressing the challenge of sound field reconstruction \cite{pezzoli2022deep, pezzoli2023implicit, fernandez2023generative, lluis2020sound,karakonstantis2024room}. 
Following the paradigm of image inpainting~\cite{liu2018image}, Lluis et al. in~\cite{lluis2020sound} propose a CNN trained on an extensive dataset of Room Transfer Functions (RTFs). This approach forces the network to learn features from RTFs under various acoustic conditions, enabling its applicability to unseen rooms with similar acoustic conditions.  However, they only consider the magnitude of the RTFs.
The majority of learning-based approaches primarily handles real-valued features~\cite{barrachina2021complex}, leaving the inference of the phase to other processing blocks leading to suboptimal results.
In this context, the adoption of Complex-Valued Neural Networks (CVNNs) is a convenient choice, due to their ability to directly handle complex-valued data and optimization~\cite{trabelsi2018deep}. This makes them well-suited to address various audio signal processing problems \cite{halimeh2021combining, watcharasupat2022end, comanducci2022synthesis
}. 

In this paper, we present a novel approach based on CVNNs and irregularly distributed microphones for addressing the RTF reconstruction problem. 
To the best of our knowledge, this is the first application of CVNNs in the context of RTF reconstruction. Additionally, the approach showcases its efficacy by successfully handling the reconstruction task with a minimal number of irregularly distributed microphones.
In particular, we consider a 2D grid of positions deployed on a plane where we aim to estimate the RTFs. The input of the CVNN is a sparse set of RTFs measured at a few points on the grid, while the outputs are the RTFs estimated at all the grid points.

We compare the proposed model with the signal processing
kernel-based approach proposed in~\cite{ueno2018kernel}, demonstrating the benefits of incorporating CVNNs into the RTF reconstruction problem. 
For informative purposes, we report evaluation results comparing the proposed model with the data-driven approach proposed in~\cite{lluis2020sound}, which reconstructs the magnitude of the sound field. In~\cite{lluis2020sound}, the authors approach the RTFs reconstruction as an image inpainting task but their method lacks the capability to recover the whole complex sound field. While acknowledging the dissimilarity in our respective objectives, we recognize the importance of providing insightful outcomes by applying the data-driven model to a similar task. 

The code to reproduce the study is publicly available on GitHub
\footnote{\href{https://github.com/RonFrancesca/complex-sound-field}{https://github.com/RonFrancesca/complex-sound-field/}}.

\section{Data Model and Problem formulation}
\label{sec:ii-probform}

In the scope of this study, we consider an acoustic source
positioned at $\mathbf{s} \in \mathbb{R}^3$ and an array of microphones deployed on a two-dimensional $W\times H$ grid in a shoebox room of dimension $L_x \times L_y \times L_z$. We define the coordinates of a point $\mathbf{r}_{w,h}$ on the grid as:
\begin{equation}
    \mathbf{r}_{w,h} = [w [L_x/(W-1)], h [L_y/(H-1)], \overline{z}]^T,
\end{equation}
where $w=0,\ldots,W-1$ and $h=0,\ldots,H-1$ are the indexes of the microphones on the grid and $\overline{z}$ is a fixed value on the z-axis. The RTF from source $\mathbf{s}$ to microphone $\mathbf{r}_{w,h}$ in a lightly damped shoebox room can be computed using an infinite summation of room modes, expressed as 
\begin{equation}
    G(\mathbf{r}_{w,h}|\mathbf{s},\omega) = -\frac{1}{V}\sum_{n}^{\infty}\frac{\Psi_{n}(\mathbf{r}_{wh})\Psi_{n}(\mathbf{s})}{(\omega/c)^2 - (\omega_{n}/c)^2 - j\omega/\tau_{n}},
\end{equation}
where $\omega$ denotes the angular frequency, $c$ the speed of sound, $\tau_n$ the decay time of the considered mode and $\Psi(\cdot)$ the corresponding mode shape. For the sake of notational simplicity, we denote the mode, identified through the 3-dimensional index $[n_x,n_y,n_z]$ with the index $n$, which describes its position on the frequency axis.
The RTFs from source $\mathbf{s}$ to points $\mathbf{r}_{w,h}$ at frequency $\omega$ are collected in $\mathbf{G} \in \mathbb{C}^{W\times H}$ such that:
\begin{equation}
    [\mathbf{G}]_{w,h}(\mathbf{s},\omega) = G(\mathbf{r}_{w,h}|\mathbf{s},\omega).
\end{equation} 
We define the set 
\begin{equation}
    \mathcal{I}_M = \{(\tilde{w}, \tilde{h})|0<\tilde{w}<W-1, 0<\tilde{h}<H-1\},
\end{equation} 
corresponding to positions on the 2D grid where microphones are not deployed. The corresponding complex-valued RTFs measured on such an incomplete grid can be then defined as:
\begin{equation}
    [\tilde{\mathbf{G}}]_{w,h}(\mathbf{s},\omega)=
    \begin{cases}
      0, & \text{if}\ (w,h) \in \mathcal{I}_M \\
      [\mathbf{G}]_{w,h}(\mathbf{s},\omega), & \text{otherwise}
    \end{cases}.
  \end{equation}
The RTF reconstruction problem is thus formulated as finding the function $\mathcal{U}(\cdot)$ that provides an estimate $\hat{\mathbf{G}}$ of the RTF matrix imposing that (omitting the frequency dependence)
\begin{equation}
    \hat{\mathbf{G}} = \arg \min_{\mathcal{U(\cdot)}} \sum\nolimits_{w=0}^{W-1}\sum\nolimits_{h=0}^{H-1}|[\mathbf{G}]_{w,h}-[\mathcal{U(\tilde{\mathbf{G}})}]_{w,h}|^2.
\end{equation}
\section{CVNN for RTF reconstruction}\label{sec:iii-method}
\begin{figure*}[!htb]%
\begin{minipage}[b]{.18\linewidth}%
  \centering%
\centerline{\includegraphics[width=.8\columnwidth]{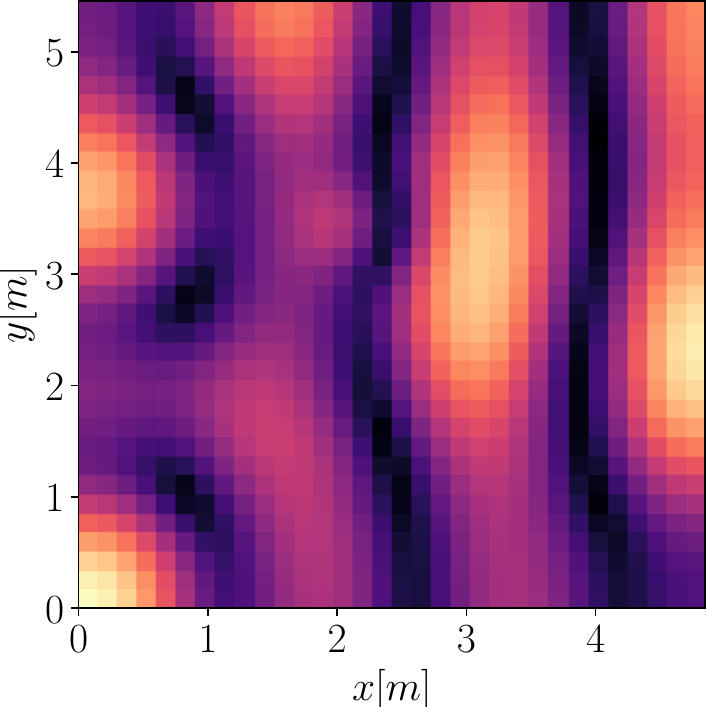}}
  \centerline{(a)}%
  \medskip%
\end{minipage}%
\hfill%
\begin{minipage}[b]{.18\linewidth}%
  \centering%
\centerline{\includegraphics[width=.8\columnwidth]{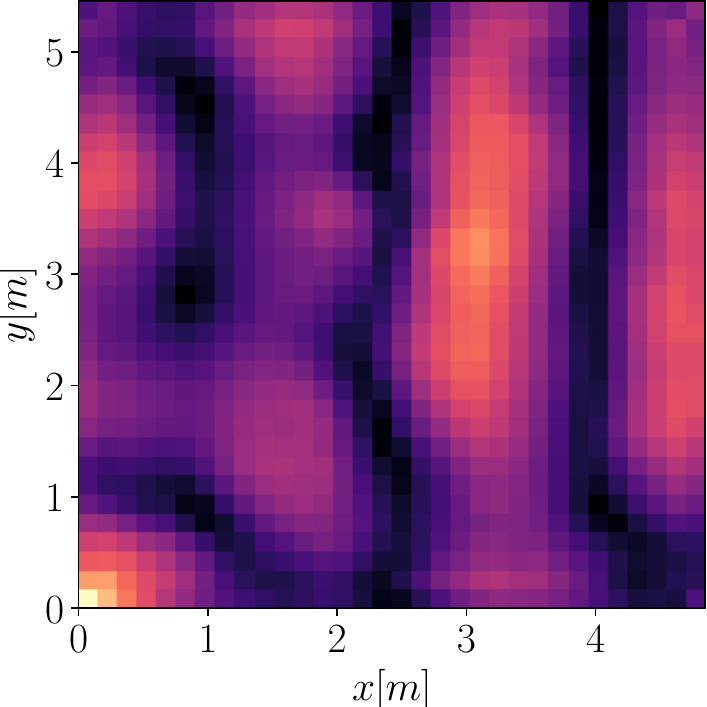}}
  \centerline{(b)}%
  \medskip%
\end{minipage}%
\hfill%
\begin{minipage}[b]{.18\linewidth}%
  \centering%
\centerline{\includegraphics[width=.8\columnwidth]{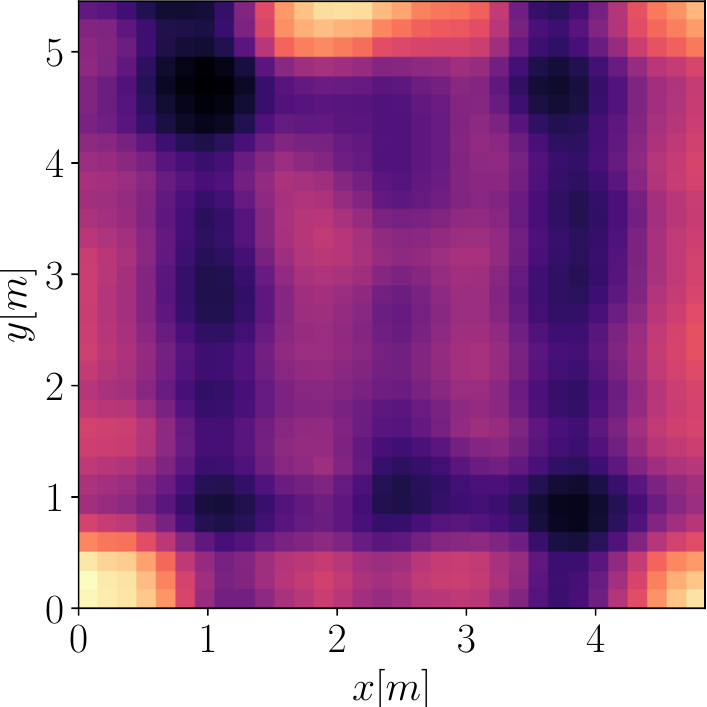}}
  \centerline{(c)}%
  \medskip%
\end{minipage}%
\hfill%
\begin{minipage}[b]{.18\linewidth}%
  \centering%
\centerline{\includegraphics[width=.8\columnwidth]{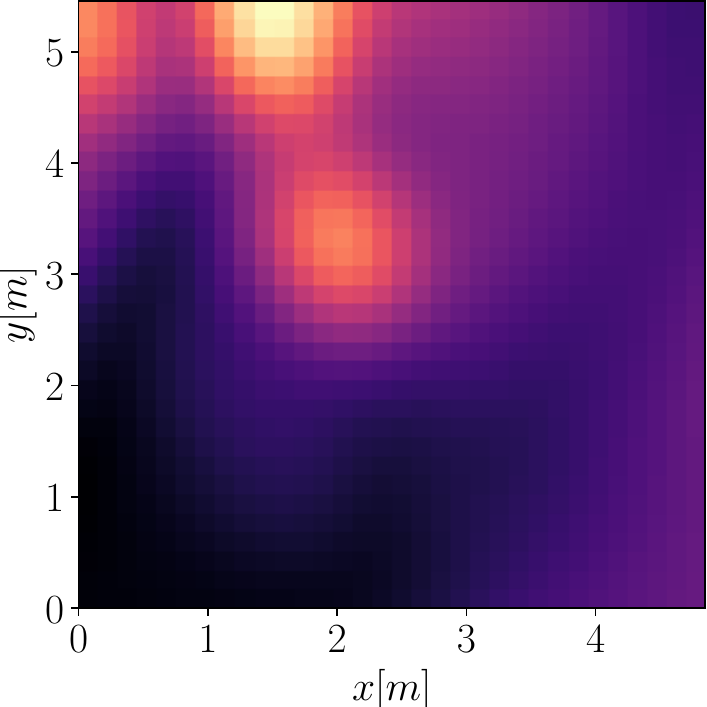}}
  \centerline{(d)}%
  \medskip%
\end{minipage}%
\hfill%
\begin{minipage}[b]{.18\linewidth}%
  \centering%
\centerline{\includegraphics[width=.8\columnwidth]{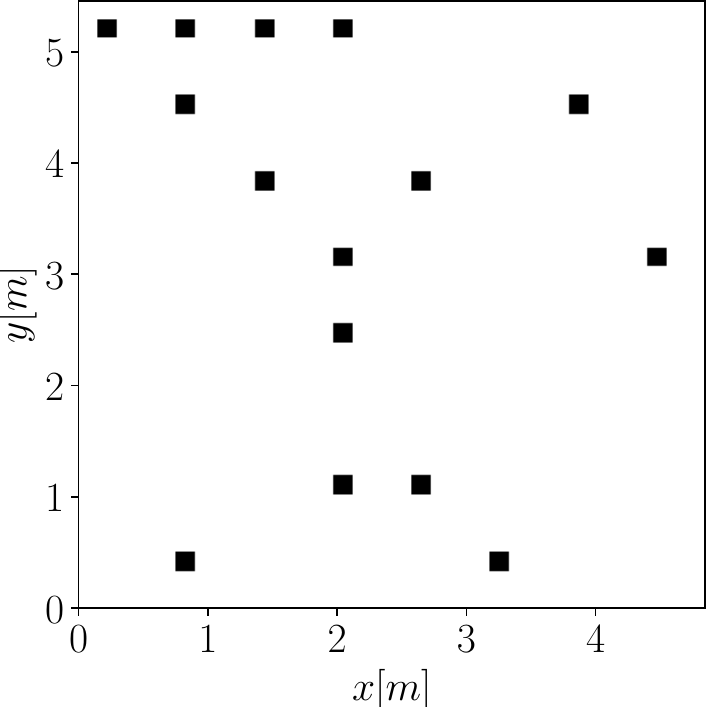}}
  \centerline{(e)}%
  \medskip%
\end{minipage}
\caption{Magnitude of the sound field, using: CVNN method (b), Lluis et al.~\cite{lluis2020sound} (c), Ueno et al.~\cite{ueno2018kernel} (d). We use the $m=15$ microphones configuration shown in (e). Ground truth magnitude shown in (a). The size of the room is $[4.8~\mathrm{m} \times 5.4~\mathrm{m} \times 2.4~\mathrm{m}]$. A source at $100~\mathrm{Hz}$ is placed at $[2.1~\mathrm{m}, 2~\mathrm{m}, 1.2~\mathrm{m}]^T$. 
}
\label{fig:example_magnitude}
\end{figure*}

\begin{figure}[!htb]
\begin{minipage}[b]{.3\columnwidth}
  \centering%
\centerline{\includegraphics[width=.9\columnwidth]{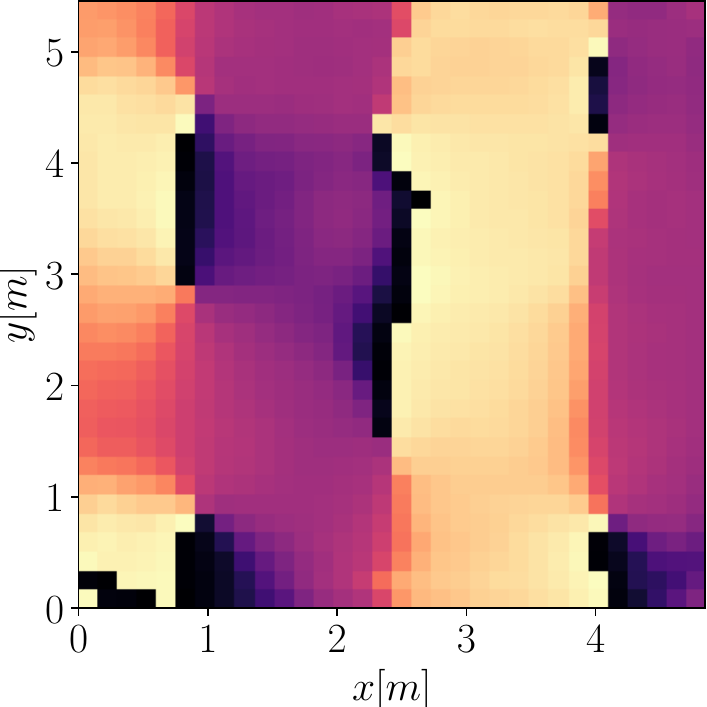}}
  \centerline{(a)}
  \medskip
\end{minipage}
\hfill
\begin{minipage}[b]{.3\columnwidth}
  \centering%
  \centerline{\includegraphics[width=.9\columnwidth]{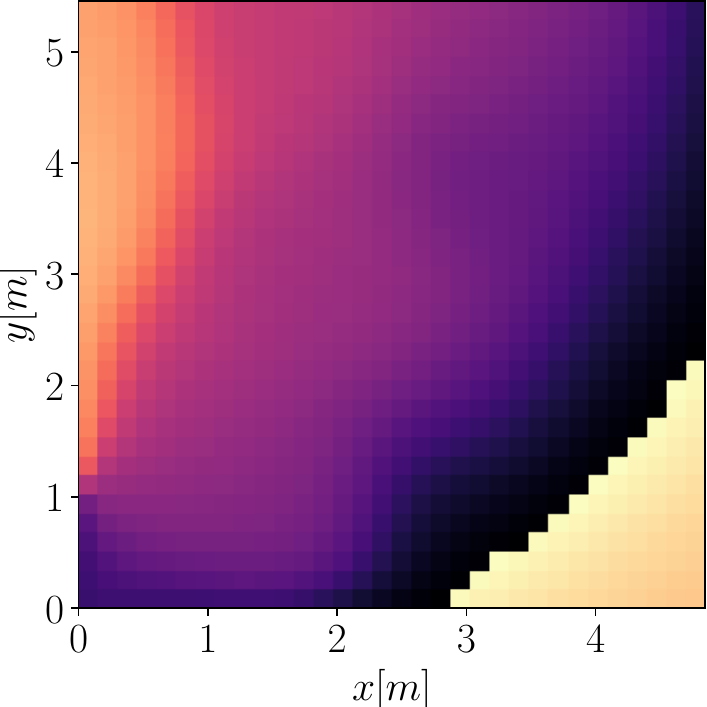}}
  \centerline{(b)}
  \medskip
\end{minipage}
\hfill
\begin{minipage}[b]{.3\columnwidth}
  \centering%
  \centerline{\includegraphics[width=.9\columnwidth]{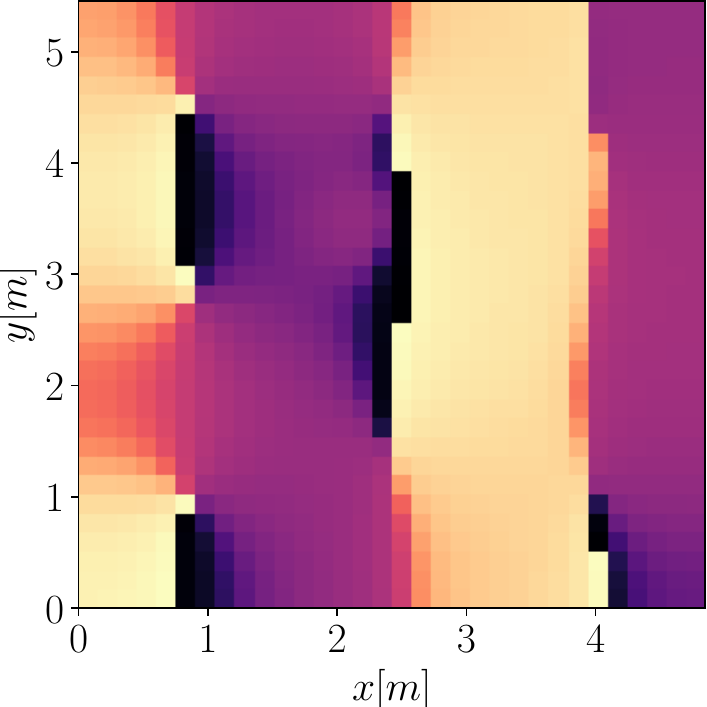}}
  \centerline{(c)}
  \medskip
\end{minipage}

\caption{Phase of the sound field obtained using the same configuration as in Fig.~\ref{fig:example_magnitude} using: CVNN (a), Ueno et al.~\cite{ueno2018kernel} (b). Ground truth shown in (c). }%
\label{fig:example_phase}
\end{figure}

In this study, we use convolutional layers considering complex-values tensors~\cite{trabelsi2018deep}, CPReLU as activation function~\cite{pandey2019exploring}, and complex-gradient to perform backpropagation~\cite{ amin2011wirtinger}. 
A theoretical analysis of the redefinition of primary operations and activation functions in the complex domain is out of the scope of the paper, and a mathematical explanation has been exhaustively given in 
~\cite{trabelsi2018deep, kuroe2003activation, pandey2019exploring}. 

\begin{figure*}%
\centering%
\begin{minipage}[b]{.28\linewidth}%
  \centering%
\centerline{\includegraphics[width=\columnwidth]{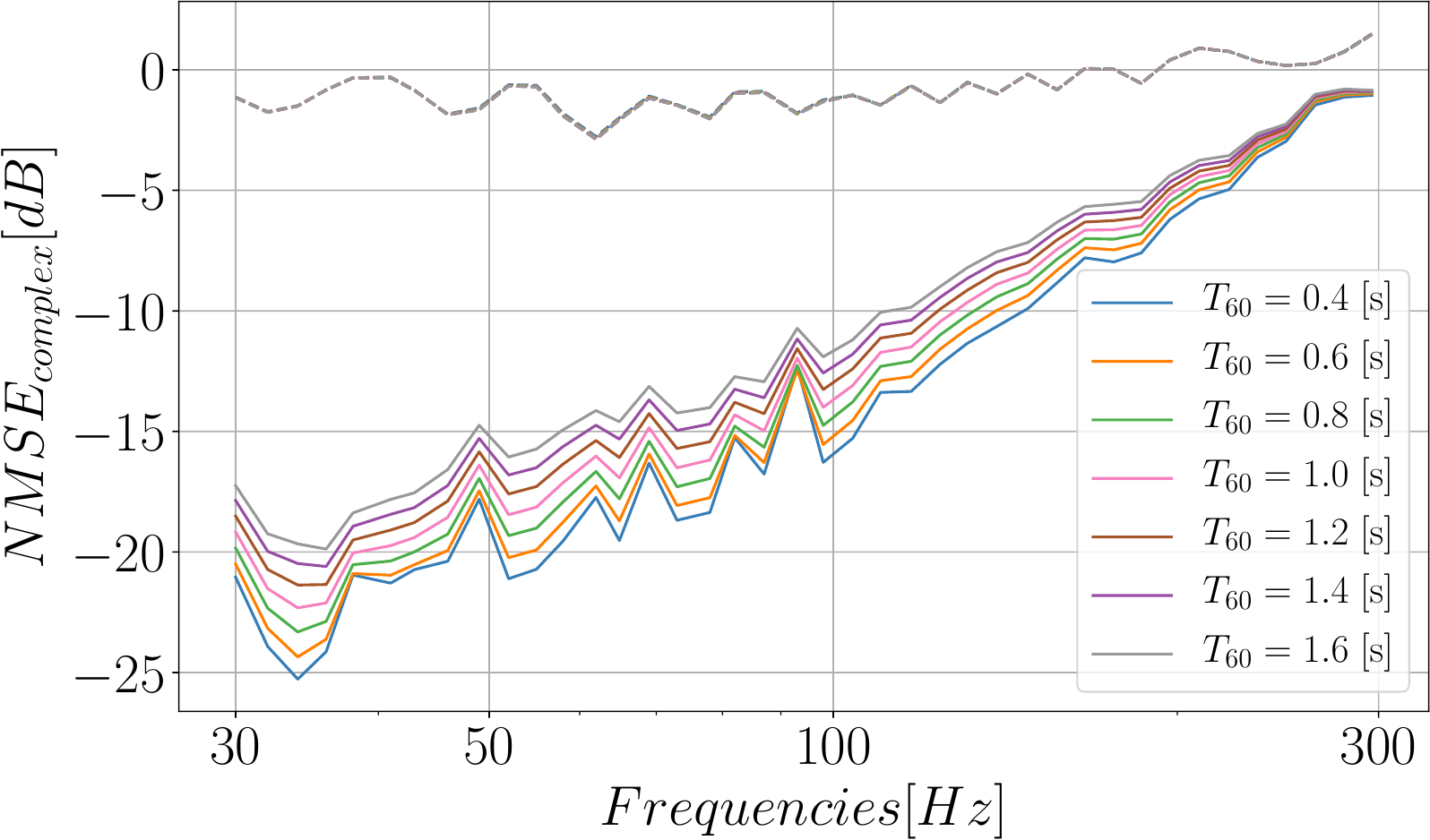}}%
  \centerline{(a)}%
  \medskip%
\end{minipage}%
\hfill%
\begin{minipage}[b]{.28\linewidth}%
  \centering%
\centerline{\includegraphics[width=\columnwidth]{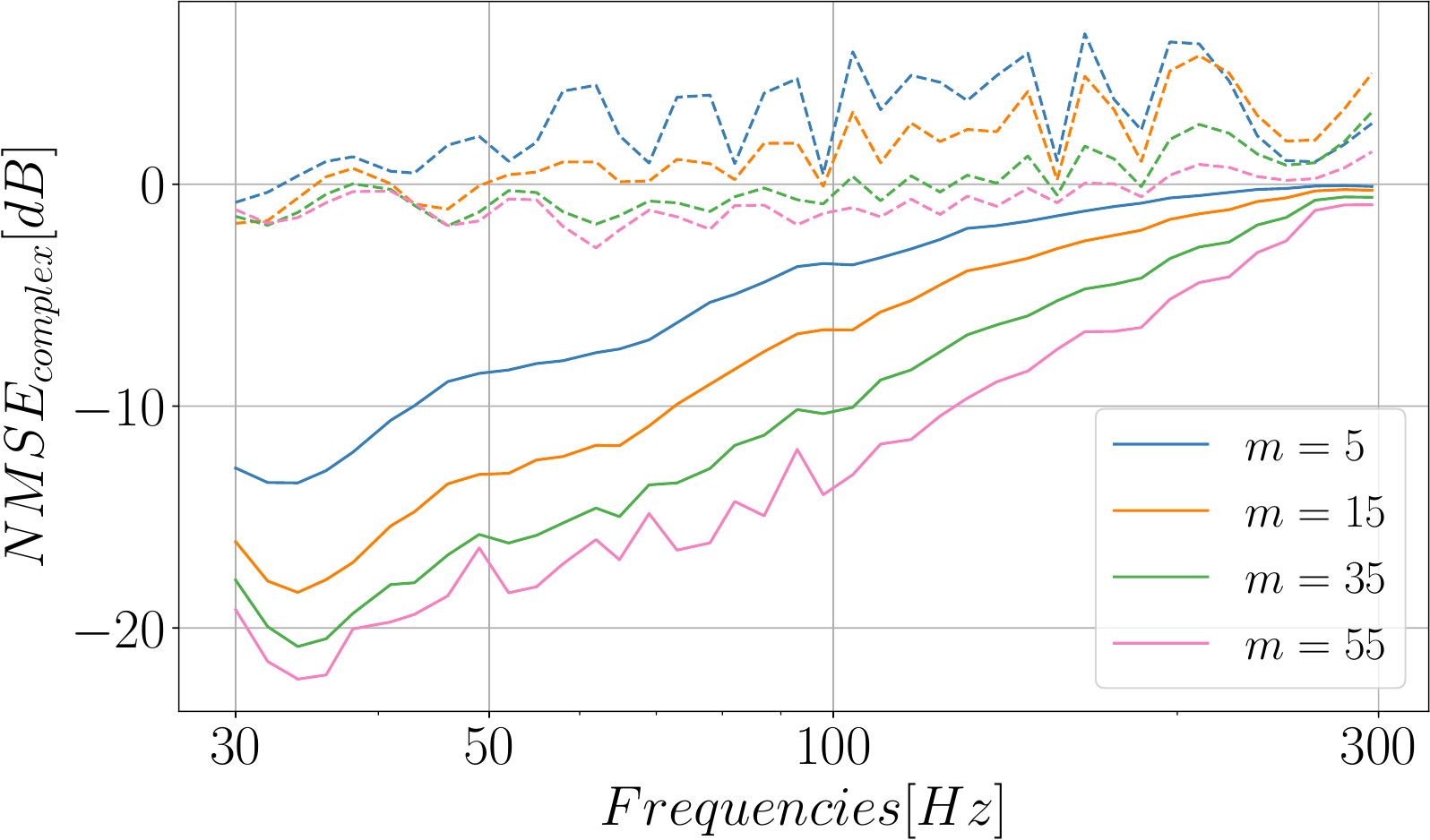}}
  \centerline{(b)}%
  \medskip%
\end{minipage}%
\hfill%
\begin{minipage}[b]{.28\linewidth}%
  \centering%
\centerline{\includegraphics[width=\columnwidth]{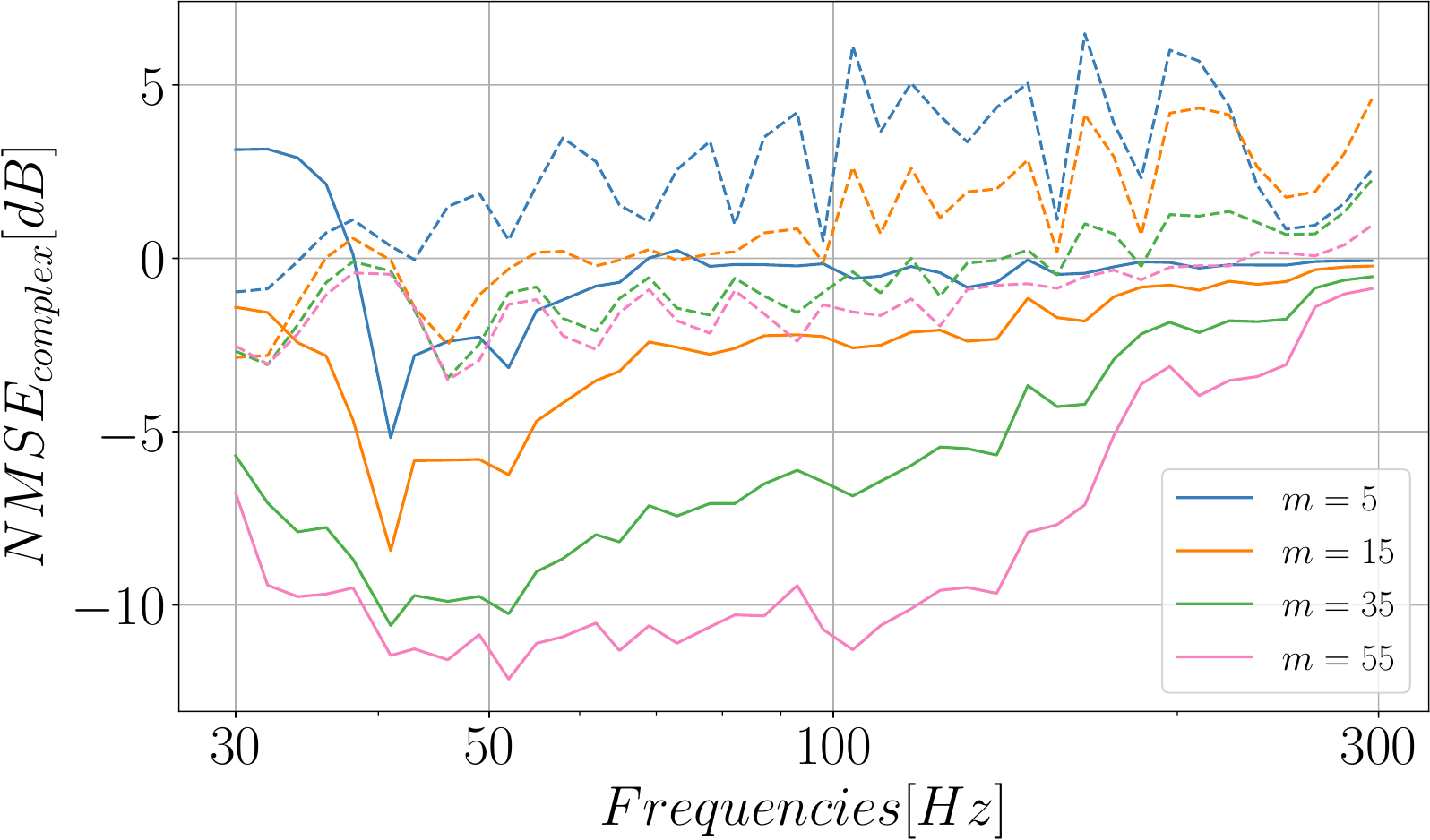}}%
  \centerline{(c)}%
  \medskip%
\end{minipage}%
\caption{(a) $\mathrm{NMSE}_\text{complex}$  calculated over simulated data with varying $T_{60}$ levels and fixed number of microphones $m=55$. (b) $\mathrm{NMSE}_\text{complex}$ calculated over simulated data with varying number of microphones $m$ and fixed $T_{60}=\SI{1}{\second}$. (c) $\mathrm{NMSE}_\text{complex}$ calculated over real data with varying number of microphones $m$ and fixed $T_{60}=\SI{1}{\second}$. The straight line corresponds to the proposed CVNN method, while the dashed line corresponds to the kernel-based technique~\cite{ueno2018kernel}.}%
\label{fig:nmse_koyama}%
\end{figure*}
\begin{figure*}
\centering%
\begin{minipage}[b]{.28\linewidth}
  \centering%
\centerline{\includegraphics[width=\columnwidth]{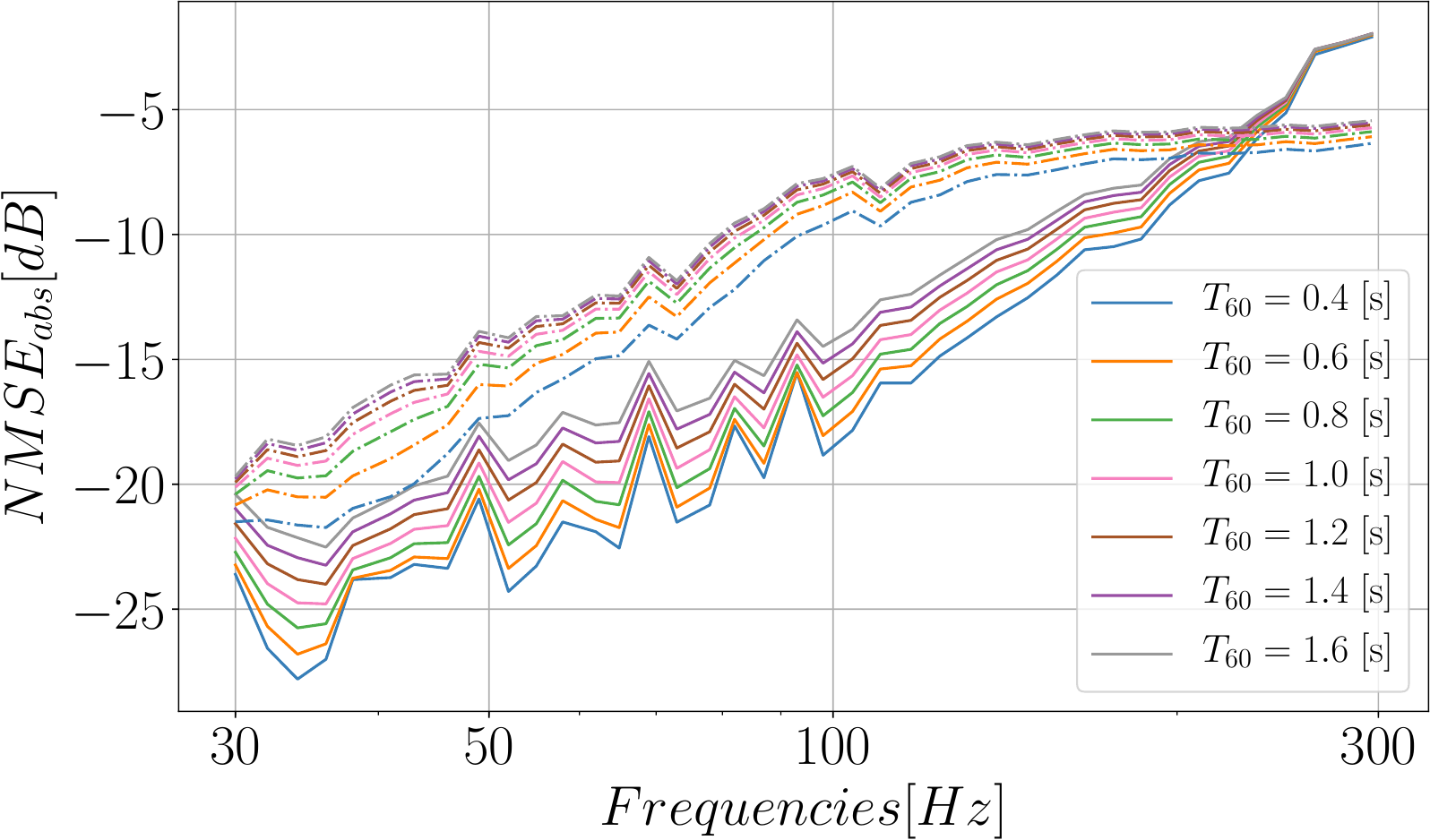}}%
  \centerline{(a)}%
  \medskip%
\end{minipage}%
\hfill%
\begin{minipage}[b]{.28\linewidth}%
  \centering%
\centerline{\includegraphics[width=\columnwidth]{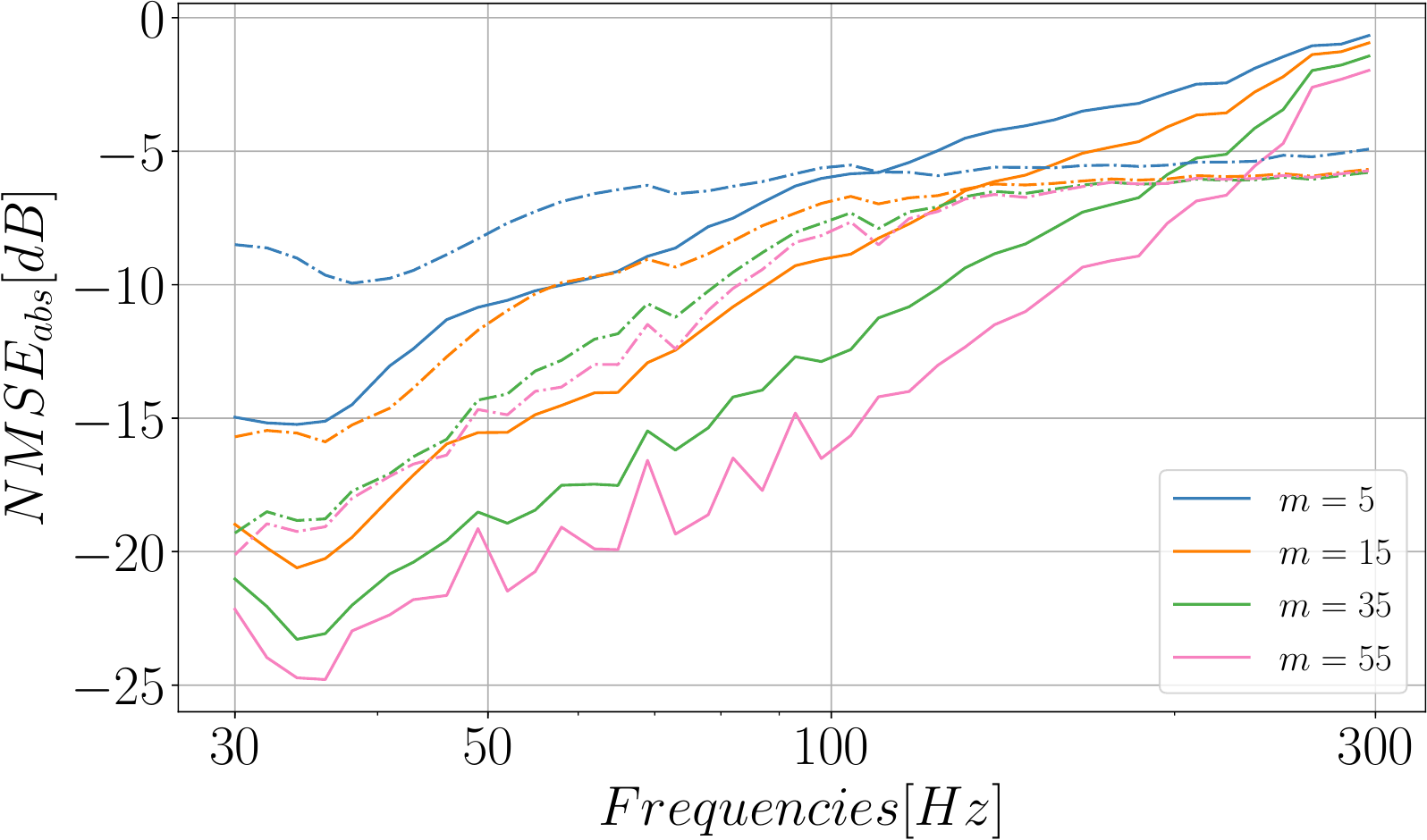}}
  \centerline{(b)}%
  \medskip%
\end{minipage}%
\hfill%
\begin{minipage}[b]{.28\linewidth}%
  \centering%
\centerline{\includegraphics[width=\columnwidth]{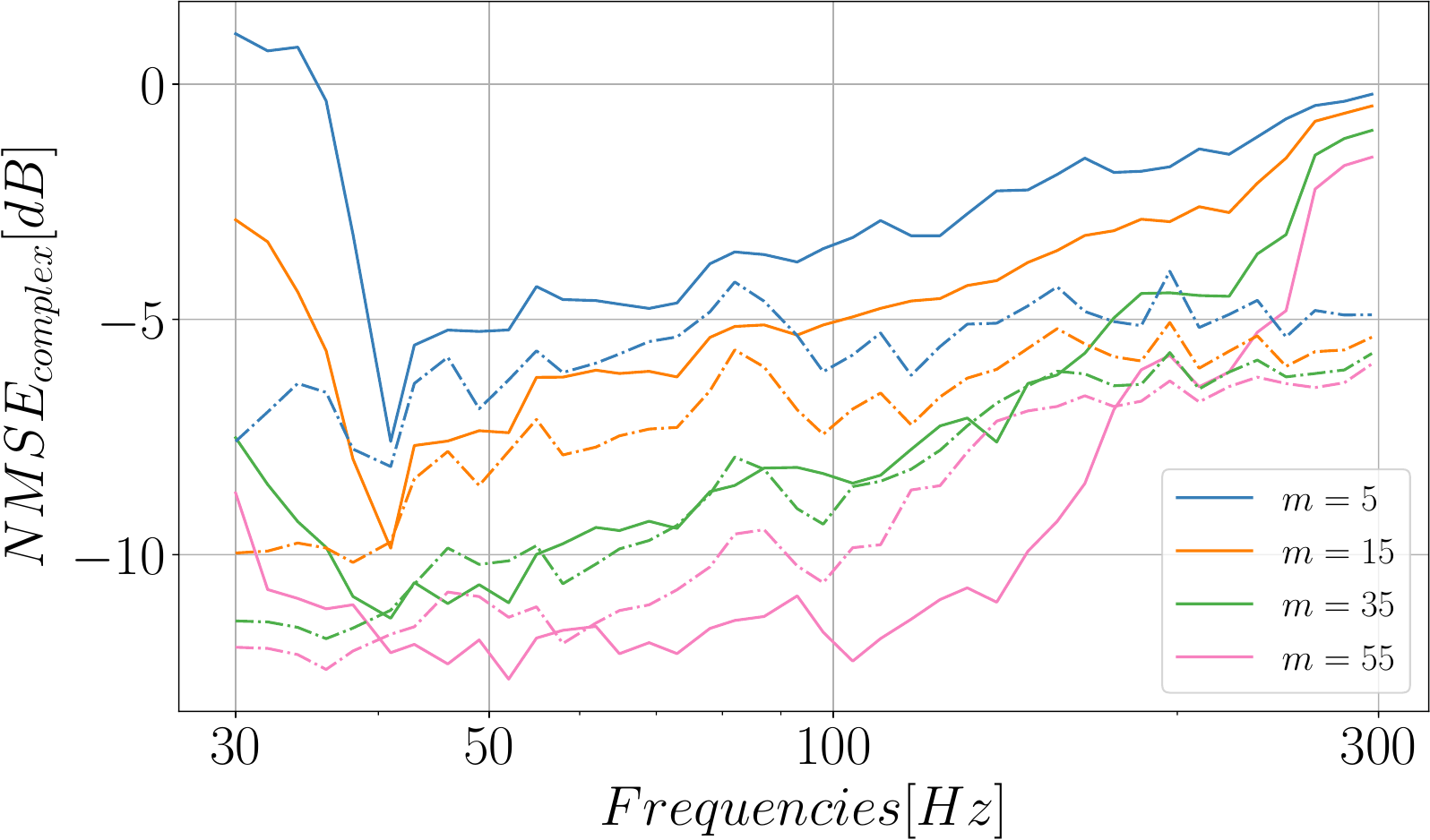}}%
  \centerline{(c)}%
  \medskip%
\end{minipage}%
\caption{(a) $\mathrm{NMSE}_\text{abs}$ calculated over simulated data with varying $T_{60}$ levels and fixed number of microphones $m=55$. (b) $\mathrm{NMSE}_\text{abs}$  calculated over simulated data with varying number of microphones $m$ and fixed $T_{60}=1~\mathrm{s}$. (c) $\mathrm{NMSE}_\text{abs}$  calculated over real data with varying number of microphones $m$ and fixed $T_{60}=1~\mathrm{s}$.  The straight line corresponds to the proposed CVNN method, while the dashed line corresponds to the data-driven approach~\cite{lluis2020sound}.}
\label{fig:nmse_lluis}
\end{figure*}
%
\subsection{Input representation}
To properly condition the proposed network model on reliable measurement positions, we concatenate the incomplete RTF matrix $\tilde{\mathbf{G}}$ with a conditioning binary mask $\mathbf{M}\in \mathbb{Z}_{2}^{W\times H}$ to create the input. 
The entries of $\mathbf{M}$ are equal to $1$ for the $w,h$ positions where we have measurements of the sound field, while 0 for all other positions. Its role is to condition the learning to focus on the points for which we have actual measurements.
Subsequently, denoting $K$ as the number of frequencies considered, the network's input consists of the matrix $\mathbf{G}_\text{input} \in \mathbb{C}^{W \times H \times 2K}$ obtained as: $\tilde{\mathbf{G}}_\text{input} = [\tilde{\mathbf{G}}~~\mathbf{M}]$
where $[\cdot]$ denotes tensor concatenation along the frequency axis.

\subsection{Network Architecture}
The CVNN proposed in this study adopts a U-Net-like network architecture~\cite{ronneberger2015u}. Specifically, the encoder is composed of four complex-valued convolutional layers with filter counts of i) 128, ii) 256, iii) 512, and iv) 1024. The decoder consists of five complex-valued convolutional layers with filter counts of v) 512, vi) 256, vii) 128, viii) 80, and ix) 80.
All encoder layers employ a stride of $2\times2$. In contrast, all decoder layers, except layer ix), are preceded by a complex-valued upsampling operation with a factor of $2\times2$. 
Each complex-valued convolutional layer is followed by CPReLU and complex-batch normalization~\cite{trabelsi2018deep}, except layers i), viii), and ix). The kernel size for all convolutional layers is $3 \times 3$, except layer ix), which has a kernel size of $1 \times 1$. Four skip connections are implemented by concatenating the inputs of layers v), vi), vii), and viii) with the inputs of layers iv), iii), ii), and i), respectively. A schematic representation of the proposed architecture is shown in Fig.~\ref{fig:architecture}.

\subsection{Training Procedure}
In the training phase, the matrix $\tilde{\mathbf{G}}_\text{input}$ is the input to the CVNN $\mathcal{U}(\cdot)$. This network produces an estimate $\hat{\mathbf{G}}$ of the ground truth complex-valued RTFs $\mathbf{G}$. 
The loss $\mathcal{L}: \mathbb{C} \rightarrow \mathbb{R}$ employed for backpropagation is computed as the $\ell_1$ norm, expressed as
\begin{equation}
    \mathcal{L}(\mathbf{G},\hat{\mathbf{G}}) = \sum\nolimits_{w=0}^{W-1}\sum\nolimits_{h=0}^{H-1}|[\mathbf{G}]_{w,h} -[\hat{\mathbf{G}}]_{w,h}|, 
\end{equation}
where, for simplicity, we omit the batch index, frequency dependence, and active source $\mathbf{s}$.

\section{EXPERIMENTAL VALIDATION}\label{sec:iv-results}

We compare the proposed method with the signal processing kernel-based interpolation technique proposed by Ueno et al. in~\cite{ueno2018kernel}.
To show the effectiveness of the proposed system, we compare the two approaches on both simulated and real data. 
For informative purposes, we also report evaluation results comparing the proposed CVNN with the data-driven approach proposed in~\cite{lluis2020sound}. We would like to stress the fact that the two methods do not address exactly the same task, since in~\cite{lluis2020sound} only the magnitude of the sound field is reconstructed. 
We show an example of reconstructed sound field for a single source and all considered methods in Fig.~\ref{fig:example_magnitude} and Fig.~\ref{fig:example_phase}, for what concerns the magnitude and phase, respectively.

\subsection{Setup}
To train the network-based models, we generated a dataset of $5000$ rooms, divided with a 75/25\% split between the training and validation sets, respectively. The generation process followed the methodology outlined in~\cite{lluis2020sound}. 
Room dimensions were selected following the ITU-R BS.1116-3 standard for listening rooms, as referenced in \cite{lluis2020sound}.
The $T_{60}$ of each room was randomly selected from a range between $\SI{0.4}{\second}$ and $\SI{1.6}{\second}$, with intervals of $\SI{0.2}{\second}$. 

For the simulation data, a distinct evaluation dataset for each $T_{60}$ considered at training has been generated. Each evaluation dataset contains $15.000$ simulated rooms, generated using the same dimensions procedure as the training dataset. To test the models on real data, we considered \textit{Room B} of the ISOBEL dataset proposed by Kristoffersen et al. in~\cite{kristoffersen2021deep}. Room B has dimensions $[\SI{4.16}{\meter}\times \SI{6.46}{\meter} \times \SI{2.30}{\meter}]$ and contains $2$ sources. We generate $100$ microphone configurations for each source. Reverberation time is around $T_{60}=\SI{1.17}{\second}$. For further details regarding the ISOBEL dataset, the reader is referred to \cite{kristoffersen2021deep}.

During the training of the networks, for each room, the number of microphones was randomly selected from a set of $5$, $10$, $15$, $35$, and $55$ microphones. The microphones were then irregularly distributed in the considered room. 

The proposed CVNN has been trained for a maximum of $5000$ epochs using a mini-batch of size $32$ and the Adam optimizer with the default configuration and a learning rate of $0.001$.
Lluis et al. model~\cite{lluis2020sound} was trained for a maximum of $5000$ epochs using a mini-batch size of $4$ (as proposed in the original paper), and the Adam optimizer with a learning rate $0.0001$. Both methods implemented an early stop condition with patience on the validation loss of $100$ epochs. 
Ueno et al.~\cite{ueno2018kernel} technique was used with default parameters and a regularization factor of $0.01$. We considered $K=40$ frequencies in the modal frequency range~\cite{schmid2021spatial} between $\SI{30}{\hertz}$ and $\SI{300}{\hertz}$. The considered range of [30, 300]~$\si{\hertz}$ includes all the room modes with a resonance frequency below $\SI{400}{\hertz}$, which gives K = 40 frequency points~\cite{lluis2020sound}. 
\subsection{Evaluation Metric}
The models have been evaluated using two Normalized Mean Squared Error (NMSE) metrics. When considering only the magnitude of the RTFs, the we define:
{\footnotesize\begin{equation}
    \mathrm{NMSE}_\mathrm{abs} = 10\log_{10} \frac{\sum_{w=0}^{W-1}\sum_{h=0}^{H-1} ||[\hat{\mathbf{G}}]_{w,h}|- |[\mathbf{G}]_{w,h}| |^2}{\sum_{w=0}^{W-1}\sum_{h=0}^{H-1} |[\mathbf{G}]_{w,h}|^2}.
\end{equation}}
Considering the entire complex field, instead, we define:
{\footnotesize\begin{equation}
    \mathrm{NMSE}_\mathrm{complex} = 10\log_{10} \frac{\sum_{w=0}^{W-1}\sum_{h=0}^{H-1} |[\hat{\mathbf{G}}]_{w,h}- [\mathbf{G}]_{w,h} |^2}{\sum_{w=0}^{W-1}\sum_{h=0}^{H-1} |[\mathbf{G}]_{w,h}|^2}.
\end{equation}}

\subsection{Results on simulated and real data}
\label{subsec:sim_data}

This section reports the results when comparing the proposed CVNN and the kernel-based techniques proposed in~\cite{ueno2018kernel}. 
We here compare the two methods in two scenarios: fixing the number of the microphone at $m=55$ while varying the $T_{60}$, and fixing $T_{60}=\SI{1}{\second}$ (the average value of the ones considered) while varying the number of microphones. 

Fig.~\ref{fig:nmse_koyama}~(a) shows the complex NMSE calculated over simulated data with varying $T_{60}$ levels and fixed number of microphones $m=55$. When compared with~\cite{ueno2018kernel}, the proposed CVNN consistently overcomes the kernel-based processing method. As expected, the higher the $T_{60}$, the more the error increases, but the proposed method never reaches $\SI{0}{\decibel}$, in contrast to~\cite{ueno2018kernel}. 

Fig.~\ref{fig:nmse_koyama}~(b) illustrates the complex NMSE calculated over simulated data with a varying number of microphones $m$ and fixed $T_{60}=\SI{1}{\second}$. The results confirm the conclusion given for Fig.~\ref{fig:nmse_koyama}~(a). Also for this case, the proposed CVNN method exceeds the performance of the kernel-based method. The error decreases as the number of microphones increases, likely due to the greater amount of input information. Overall, the results show that employing a complex-valued neural network improves the reconstruction of a RTF. This is primarily attributed to the neural network's superior capability compared to a kernel-based interpolation method, enabling enhanced generalization across a variety of environmental conditions and a wider spatial area. 
The major gain in performance is at the lower frequencies. This is probably due to the fact that at high frequencies, the spatial variability of the modal patterns is more accentuated for both absolute value and phase of the sound field, making the reconstruction process more challenging. The understanding of how to enhance this issue requires further investigations that will be carried out in future works.

Fig.~\ref{fig:nmse_koyama}~(c) shows the complex NMSE calculated over real data with a varying number of microphones $m$. The $T_{60}$ is the one of the real ISOBEL room (see Section \ref{sec:iv-results} for further details). 
The results confirm what has been already discussed in Section \ref{subsec:sim_data}. 
The proposed CVNN model overcomes the kernel-based signal processing method in all the different configurations considered, also in the case of real data. 
The higher the number of microphones, the higher the gain of the proposed method for the same conclusion given for the simulated data. 
Overall, the results show that the proposed CVNN performs better than kernel-based interpolation for reconstructing RTF over a wide region. 
\subsection{Comparison with magnitude-inpainting-techniques}
In this section, we report results comparing the proposed model with the data-driven approach proposed in~\cite{lluis2020sound}. 
Differently from the proposed CVNN method, in~\cite{lluis2020sound} the RTFs reconstruction task is approached as an image-inpainting task, where only the magnitude of the sound field is considered. 
Therefore, for our method as well, we report the results considering only the $\mathrm{NMSE}_\text{abs}$. 

Fig.~\ref{fig:nmse_lluis}~(a) shows the $\mathrm{NMSE}_\text{abs}$  calculated with varying $T_{60}$ levels and fixed microphones $m=55$, while Fig.~\ref{fig:nmse_lluis}~(b) and Fig.~\ref{fig:nmse_lluis}~(c) reports the $\mathrm{NMSE}_\text{abs}$  calculated over a varying number of microphones $m$ and fixed $T_{60}=\SI{1}{\second}$, on simulated and real data respectively. When comparing the methods for each selected number of microphones, the proposed CVNN-based method obtains better performance at lower frequencies, whereas~\cite{lluis2020sound} obtains higher performances at higher frequencies. However, the advantage at high frequencies decreases as the considered number of microphones increases. For $m=55$, the proposed method outperforms the alternative method across almost the entire frequency range. This behavior can be attributed to both the selected loss function, which tends to underperform at higher frequencies, and the observation that our proposed model extracts more information compared to~\cite{lluis2020sound}, while being evaluated on a smaller set of measurements in this case.  
More specifically, this implies that the proposed method undergoes a more intricate optimization process, effectively taking into account the complex-valued sound field, while~\cite{lluis2020sound} reconstructs missing patterns in the  magnitude, treating the problem as an image-inpainting one. Also in this case, the error decreases as the number of microphones increases, confirming the conclusions drawn above.
\section{Conclusions and future work}\label{sec:v-conclusion}
This paper proposed a novel application of complex-valued neural networks to the task of room transfer function reconstruction, considering a significantly small set of irregularly placed microphones, overcoming the need for an unpractical number of microphones. 
The proposed CVNN is compared with a kernel-based interpolation method, able to reconstruct the phase of the RTFs, both on simulated and real data, outperforming the latter both at low and high frequencies and generalizing on shoebox rooms. We also compare the magnitude reconstruction performance of the proposed model with a state-of-the-art data-driven technique, showing better performances at low frequencies while underperforming at higher frequencies, although the gap becomes smaller as the number of microphones increases. The results obtained motivate deeper investigation into incorporating CVNN in addressing RTF reconstruction challenges. Future work aims to enhance the performance of the proposed model by considering more suitable loss functions, including direct considerations of phase handling. Additionally, future work will focus on analyzing the behavior of the proposed model across frequency ranges beyond the modal range and arbitrary room shapes. 
\bibliographystyle{ieeetran}
\bibliography{refs}

\end{document}